\documentclass[useAMS,usenatbib,paper]{mn2e}
\usepackage{times,epsfig,psfig,natbib,amssymb,amsmath,graphicx}
\usepackage[bookmarks,bookmarksnumbered,colorlinks=true, citecolor=blue, linkcolor=black]{hyperref}
\usepackage{color,natbib}
\usepackage[normalem]{ulem}

\title[Discovery of a FR~0 radio galaxy emitting at $\gamma$-ray
  energies]{Discovery of a FR~0 radio galaxy emitting at $\gamma$-ray
  energies} \author[Paola Grandi, Alessandro Capetti ,
  Ranieri D. Baldi]{Paola Grandi$^{1}$\thanks{E-mail:grandi@iasfbo.inaf.it} ,
  Alessandro Capetti$^{2}$\thanks{capetti@oato.inaf.it} and
  Ranieri~D. Baldi$^{3}$\thanks{r.baldi@soton.ac.uk}\\ $^{1}$INAF-IASFBO, Via Gobetti 101, I-40129 Bologna,
  Italy\\ $^{2}$INAF-Osservatorio Astrofisico di Torino, Strada Osservatorio
  20, I-10025, Pino Torinese, Italy\\ $^{3}$Department of Physics and Astronomy, The University, Southampton SO17 1BJ, UK}
  
\begin{document}



\maketitle

\label{firstpage}

\begin{abstract}

We present supporting evidence for the first association of a {\it Fermi} source,
3FGLJ1330.0-3818, with the FR~0 radio galaxy Tol1326-379. FR~0s represent the
majority of the local radio loud AGN population but their nature is still
unclear. They share the same properties of FR~Is from the point of view of the
nuclear and host properties, but they show a large deficit of extended radio
emission. Here we show that FR~0s can emit photons at very high energies.
Tol1326-379 has a GeV luminosity of $L_{>1~{\rm GeV}} \sim 2\times10^{42}$ erg
s$^{-1}$, typical of FR~Is, but with a steeper $\gamma$-ray spectrum
($\Gamma=2.78\pm 0.14$). This could be related to the intrinsic jet properties
but also to a different viewing angle.

\end{abstract}


\section{Introduction}

Radio Galaxies (RGs) has been  historically divided into faint edge-darkened FR~I
and bright edge-brightened FR~II \citep{fanaroff74} on the basis of their
extended radio morphology with the transition occurring at, approximately, a
radio power of P$_{\rm 178~MHz} \sim$ 10$^{25}$ W~Hz$^{-1}$~ sr$^{-1}$. From the
optical point of view, radio galaxies are split into Broad Line Radio Galaxies
(BLRGs) and Narrow Line Radio Galaxies (NLRGs). NLRGs can be further
classified as High Excitation Galaxies (HEGs) and Low Excitation Galaxies
(LEGs) based on their optical emission line ratios \citep{buttiglione10}.  While
HEGs and BLRGs show almost exclusively a FR~II morphology, LEGs can assume
both radio morphologies.  It is believed that in LEGs, the AGN luminosity is
sustained by hot gas via  advection-dominated flow-like/Bondi accretion \citep{balmaverde08}, while
in BLRGs and HEGs by a cold geometrically--thin, optically--thick disk
\citep{best12,heckman14}.  Steep Spectrum Radio Quasars (SSRQs) are similar to
BLRGs but more distant and luminous.  FRIs and FR~IIs are considered the
parent population of BL LACs (BLs) and Flat Spectrum Radio Quasars (FSRQs),
respectively \citep{urry95}.

Being observed at large angles , the jets of RGs and SSRQs (collectively named
Misaligned AGN: MAGN) do not benefit from the strong Doppler amplification
typical of blazars. Although geometrically disfavored (i.e. less amplified),
RGs and SSRQs have been detected above 100 MeV.  The {\it Fermi} satellite
found 11 MAGN in only 15 months of Large Area Telescope (LAT) survey \citep{abdo10-MAGN} with four
of them also detected in the TeV band.  Although their number is a tiny
fraction of the total {\it Fermi} detected sources, their discovery had a
strong impact on the study of the high energy process in AGN.

A clear link, in at least two BLRGs, 3C~111 and 3C~120, has been established 
between the expulsion of bright superluminal knots from the radio core and
intense $\gamma$-ray flares \citep{grandi12,casadio15}.

The firm detection of GeV emission from the radio lobes in nearby RG
Centaurus A has shown that extranuclear extended regions can be a source of
gamma-ray photons, implying the presence of highly energetic particles at large distances from
the nucleus \citep{Abdo10-lobi}. Because their radiative lifetimes ($<1$ to 10
million years) approach plausible electron transport time scales across the
lobes, their presence is difficult to explain unless successive particle
acceleration occurs even at large distances from the black hole.

Spectral Energy Distribution (SED) studies of FR~I radio galaxies showed that
a pure, one-zone homogeneous, Synchrotron Self-Compton (SSC) emitting region
is inadequate in reproducing the radio to TeV data, stimulating the elaboration
of more complex models.  Stratified jets with different regions interacting
with each other \citep{georganopoulos03, ghisellini05,bottcher10} as well as
magnetic reconnection events along the jet \citep{Giannios10} or in the
vicinity of the black hole \citep{Khiali15} have been suggested as possible
sources of $\gamma$-ray photons. Hadronic models based on proton-photon
interaction (see for a review \citealt{bottcher12}) have also been explored
providing possible connections among AGN, Ultra High Energy Cosmic Rays and
neutrinos \citep{becker09}

All these results are based on the study of the brightest MAGN
discovered in the first years of the {\it Fermi-LAT} activity, i.e.
bright radio sources with flux density of 2Jy or more.  The recent
publication of the new {\it Fermi}-LAT point source catalog (3FGL -
\citealt{acero15}) and the {\it Fermi}-LAT AGN catalogs (3LAC -
\citealt{Ackermann-3LAC}) collecting 4 years of data now allow us to
extend the study of GeV radio galaxies to lower fluxes.  Although the
'standard' picture, i.e. the predominance of bright FR~Is among MAGN,
is confirmed, we can now enter into the unexplored territory of fainter radio
AGN.  Indeed we propose here the first association between a
3LAC gamma-ray source and a FR~0 galaxy.

The cross-match of radio and optical data favored by the advent of large area surveys has surprisingly revealed that  the bulk of the radio-galaxy population lacks prominent extended
radio structures.  \cite{best12} built a sample of radio-galaxies by
cross-correlating the Sloan Digital Sky Survey (SDSS), the National Radio Astronomy Observatory (NRAO) Very Large Array (VLA) Sky Survey (NVSS), and the 
the Faint Images of the Radio Sky at Twenty centimetres (FIRST) survey  datasets. This sample is selected at $F_{\rm 1.4} > 5$ mJy and it includes RGs
up to $z\sim$ 0.3, covering the range $L_{\rm 1.4} \sim 10^{22} - 10^{26}$ W
Hz$^{-1}$. Most of them ($\sim$
80 \%) are unresolved or barely resolved at the 5$^{\prime\prime}$ FIRST
resolution, corresponding to a limit to their size of $\sim$10 kpc \citep{baldi09}. The lack of extended radio structures, that characterize the
morphology of the traditional Fanaroff-Riley classes (FR~I and FR~II),
suggests defining these objects as FR~0s (see \citealt{baldi15b}).

FR~0s share the same properties of FR~Is from the point of view of the nuclear and host properties.
By comparing FR~Is and FR~0s of similar AGN power, estimated from the optical line luminosity, FR~0s show the same radio core power but a strong deficit of extended radio emission. Most of them ($\sim 70 \%$) can be classified as LEGS, the same spectroscopic class to which FR~Is belong.
Finally, the hosts of these two classes are effectively indistinguishable being in both cases red early-type galaxies with central black hole masses larger than
$\sim 10^8 M_{\odot}$ \citep{baldi09,baldi10b,sadler14,baldi15}.

Although FR~0s represent the majority of the low luminosity RGs in the local Universe, they are a puzzling class of AGN completely unexplored at high energies. 

In this paper, we show that the galaxy Tol1326-379, identified as the
counterpart of the $\gamma$-ray source 3FGLJ1330.0-3818 and classified in the
3LAC as a Flat Spectrum Radio Quasar, is actually the first FR~0 radio source
discovered in the GeV sky.

A cosmology with $H_0 = 67$ km s$^{-1}$ Mpc$^{-1}$, $\Omega_m = 0.32$, and
$\Omega_\Lambda = 0.68$ is assumed.

\section{3FGLJ1330.0-3818 - Tol1326-379}

The $\gamma$-ray source 3FGLJ1330.0-3818 was detected by {\it Fermi} with a
significance of $\sim 5 \sigma$, a flux above 1 GeV of F$_{>1~{\rm GeV}}=(3.1\pm
0.8) \times 10^{-10}$ phot cm$^{-2}$ s$^{-1}$ and a power law index
$\Gamma=2.78\pm0.14$ \citep{acero15}.

3FGLJ1330.0-3818 is listed in the 3LAC catalog and associated to the early
type galaxy Tol1326-379 at z=0.02843 with a Bayesian probability of
association of 90\%.  The Bayesian method calculates the posterior probability
that a source from a catalog of candidates is the counterpart of a
$\gamma$-ray source detected by the LAT, evaluating the significance of a
spatial coincidence between the candidate counterpart and the LAT-detected
source. For more details see \cite{abdo10-1FGL}.

\begin{figure}
\centering 
\includegraphics[width=0.8\columnwidth]{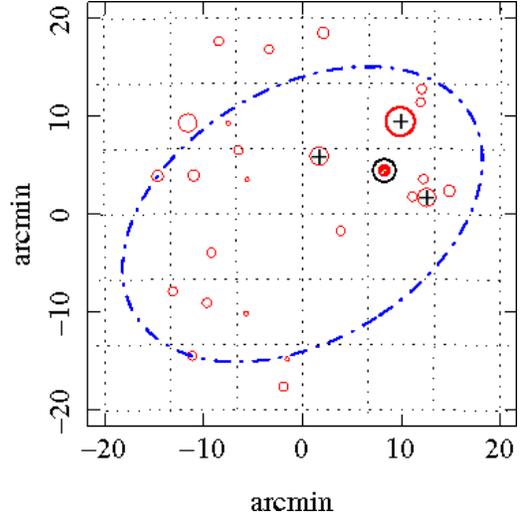}
\caption{NVSS radio sources in the field of 3FGLJ1330.0-3818. Symbol sizes are proportional to the flux density at 1.4 GHz.
Tol1326-379 is marked with the red point.  Three bright radio sources, NVSS J132911-380918, NVSS J132952-381251, and NVSS J132857-381703 (cross symbols) are within the Fermi position error ellipse, but they are rejected as possible counterparts (see text). This image was produced by using the ASDC tool http://www.asdc.asi.it/fermi3fgl/}
\label{nvss}
\end{figure}

In order to strengthen the association proposed by the 3LAC catalog, we
explored the region around 3FGLJ1330.0-3818 within the 95\% $\gamma$-ray
position error ellipse (20$\arcmin$ $\times$ 13$\arcmin$) looking for
alternative identifications. We considered all radio sources with a NVSS flux
density larger than 20 mJy ($\sim$3 times fainter than Tol1326-379), finding
three objects spanning the range 70 to 240 mJy (see Figure \ref{nvss}). Their radio spectral indices,
estimated between 0.843 and 1.4 GHz, are steep ($\alpha_{\rm r}$ $\sim$
0.7-0.9) and none of them is detected at 4.8 GHz with the PARKES telescope
\citep{griffith09}, implying that their flux densities are less than $\sim$35
mJy, suggesting the dominance of the extended radio emission. Furthermore none
of them has a 2MASS counterpart apart from Tol1326-379 which shows an
early-type morphology. This confirms that Tol1326-379 is the most likely
association.

\begin{figure*}
\centering  
\includegraphics[width=1.00\columnwidth]{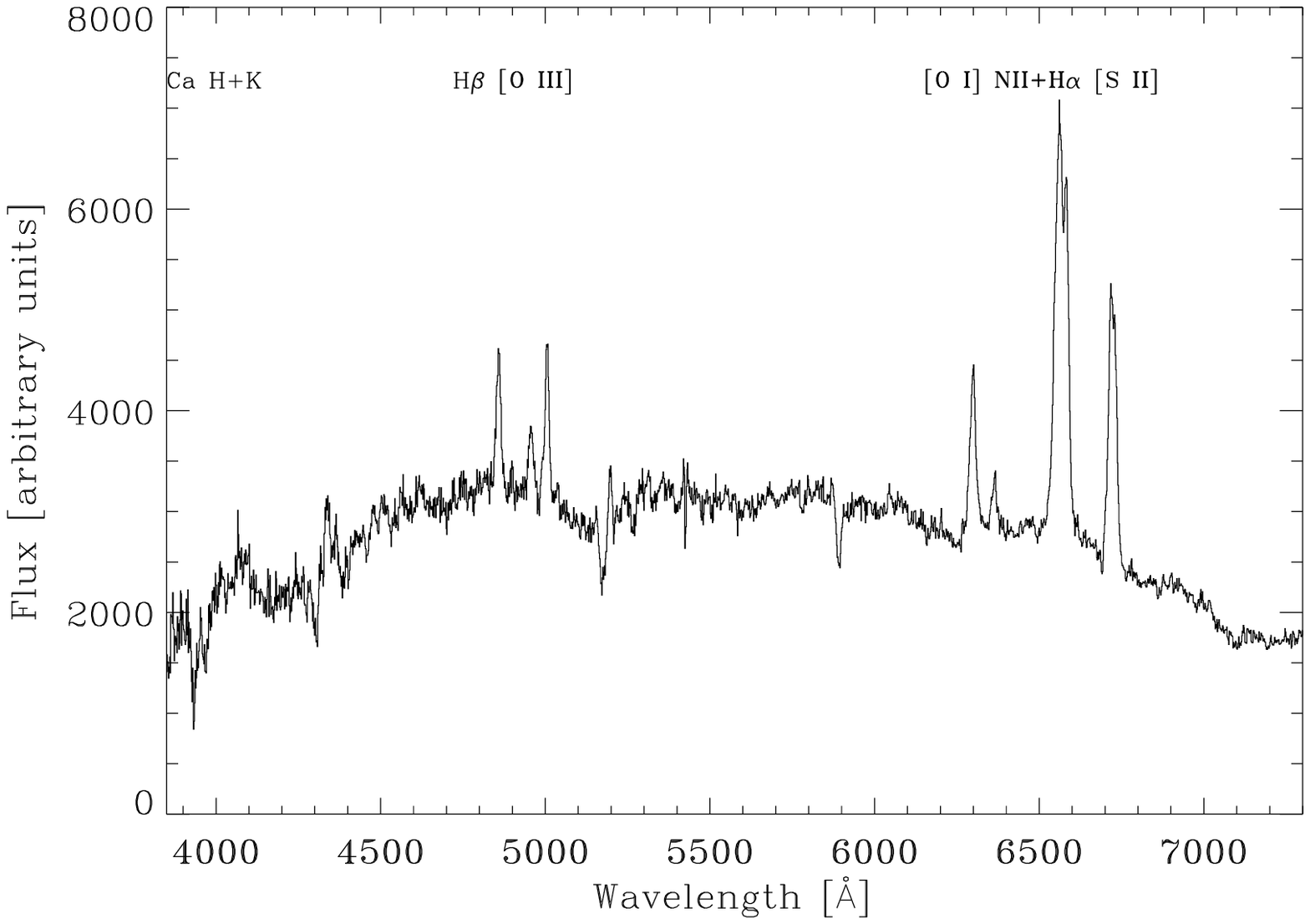}
\includegraphics[width=1.00\columnwidth]{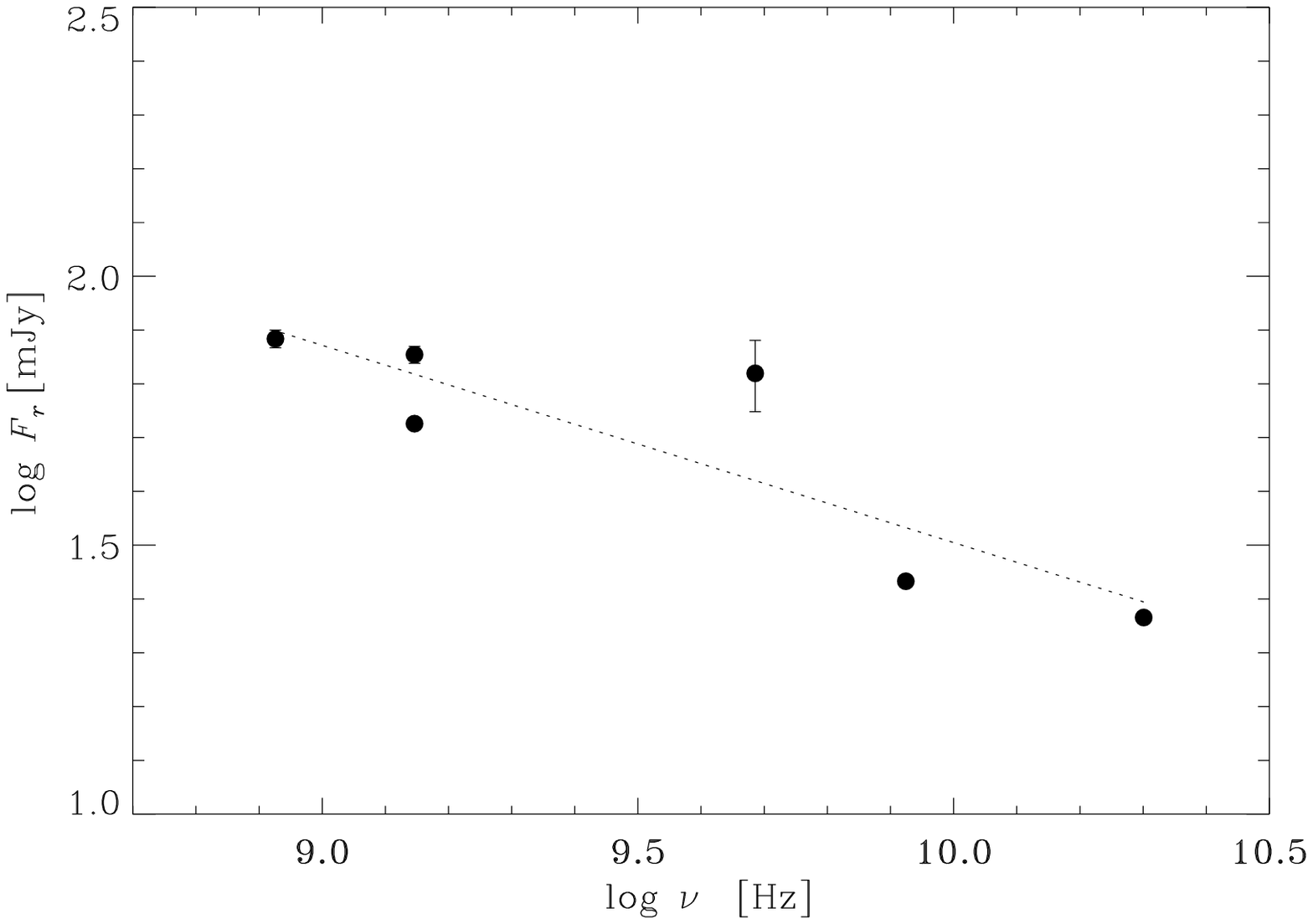}
\caption{Left: optical spectrum of Tol1326-379 from the 6dF Galaxy Survey
  \citep{jones04}. Right: radio spectrum of Tol1326-379. Data are from
  literature \citep{healey07, pmn, vla,atca, sumss}. The 20 GHz flux density was
  measured with the Australian Telescope Compact Array (see text for the
  details). The dotted line is the single power fit to the data having a slope
  of 0.37.}
\label{radio}
\end{figure*}

\section{ The galaxy Tol1326-379}

\subsection{Optical properties}
The optical spectrum of Tol1326-379 from the 6dF Galaxy Survey
\citep{jones04,jones09} is presented in Fig. \ref{radio}, left panel. It does
not show any evidence for broad emission lines, arguing against its
identification as FSRQ; on the other hand, the equivalent width (EW) of the
emission lines largely exceed the limit of 5\AA\ commonly adopted for BL~Lac
objects \citep{stickel91}. We obtained, in particular, EW(H$\alpha$) $\sim$
45$\pm$4\AA\ and EW([O~III]) $\sim$ 8$\pm$1\AA. The diagnostic line ratios (log
[O~III]/H$\beta$ =0.10, log [N~II]/H$\alpha$ =-0.03, log [O~I]/H$\alpha$
=-0.40, and log [S~II]/H$\alpha$ =0.01) are characteristic of a LEG spectrum
\citep{kewley06,buttiglione10}.  This lends further weight against its
association with a FSRQ, because such sources always show a high ionization
optical spectrum \citep{shaw12}.

Unfortunately the 6dFGS spectrum is not flux calibrated. In order to obtain the
emission lines luminosity we need to rely on an indirect estimate. We derived
the flux in the J band Two Micron All Sky Survey (2MASS) image from a synthetic aperture of 6\farcs7, the
diameter of the 6dFGS fiber, after having degraded the image resolution to
match the seeing reported in the observing log. We then adopted a V-J color of
2.43 \citep{mannucci01}, typical of early-type galaxies, and obtained of
magnitude V=15.1. From its measured EW we finally obtain an [O~III] flux of
3.2$\times10^{-15}$ erg cm$^{-2}$ s$^{-1}$ and a luminosity of 4$\times10^{40}$ erg
s$^{-1}$. 

In order to assess the accuracy of such a  procedure, we performed the same
analysis on a group of 7 early-type emission line galaxies in common between
the 6dFGS and the SDSS survey. Our line measurements agree with those provided
by the SDSS database within a factor of 4.

From its 2MASS image, we derived a total K
magnitude of 11.22, corresponding to a luminosity of 1.0$\times 10^{11}$ solar
luminosity. The tight correlation between $M_{\rm BH}$ and the near-infrared
bulge luminosity proposed by \citet{marconi03} allows us to estimate (within a
factor 2) the black hole mass of $M_{\rm BH}=2 \times 10^8$
M$_{\odot}$.

Using the relation L$_{\rm bol}$=3500$~L_{\rm [O~III]}$ measured by
\citet{heckman04}, we obtain a bolometric luminosity L$_{\rm bol}=44.1$
erg s$^{-1}$ (with an uncertainty of 0.4 dex) for this source. The
Eddington-scaled accretion rate of Tol1326-379 is $\dot L =
L_{\rm bol}/L_{\rm Edd} \sim 5 \times 10^{-3}$, a value typical of LEGs
\citep{best12}.

\subsection{Radio properties}

We collected the radio flux density data from the NASA/ IPAC Infrared
  Science Archive and Extragalactic Database (NED). To these data we added our
  own measurement at 20 GHz from the Australian Telescope Compact Array
  (ATCA); the observations were obtained on Oct 27$^{\rm th}$ 2004 (project
  C1049, PI Ekers) as part of the AT20G survey \citep{murphy10}, centered at
  19.9 GHz and with a bandwidth of 256 MHz and with a duration of 3.6
  minutes.
  We used PKS~1934 -- 638 for the primary flux calibration, 1036--52 as
  bandpass calibrator and PKS~1349-439 as phase calibrator. After
  deconvolution the source appears as unresolved (angular resolution of
  10$\arcsec$); we perform a Gaussian fit using 'jmfit' task of {\it
    AIPS} software\footnote{The NRAO Astronomical Image Processing
    System (AIPS) is a package to support the reduction and analysis
    of data taken with radio telescopes.} to estimate its flux
  density of  23.2 $\pm 0.8$ mJy.
The available radio data cover the frequency range from 843 MHz to 20 GHz
(see Fig. \ref{radio}, right panel). None of the available radio images shows
extended emission and in particular Tol1326-379 is reported by
\citet{healey07} as a point source in their CRATES catalog at 8.4 GHz with an
angular resolution of $\sim2\arcsec$ provided by the ATCA observation. The
overall spectral slope is $\alpha=0.37$ but the data points show a large
scatter. This in part due to the source variability; indeed its flux density
varied from 53.2 to 71.5 mJy in the NVSS and ATCA observations, respectively,
two measurements obtained at the same frequency and with similar spatial
resolution. In addition, the observations have rather different spatial
resolution, typically $\sim45\arcsec$ for the four lower frequencies and of a
few arcsec for the data at 8.4 and 20 GHz. The radio spectrum measured between
these two high frequency (and high resolution) measurements is even flatter than the overall value,
$\alpha_{8.4-20~{\rm GHz}}=0.11$. The core dominance R parameter defined as the ratio
between the 4.85 GHz and 1.4 GHz flux densities is $\log R = -0.03$.

\medskip

\begin{figure}
\centering 
\includegraphics[width=1.0\columnwidth]{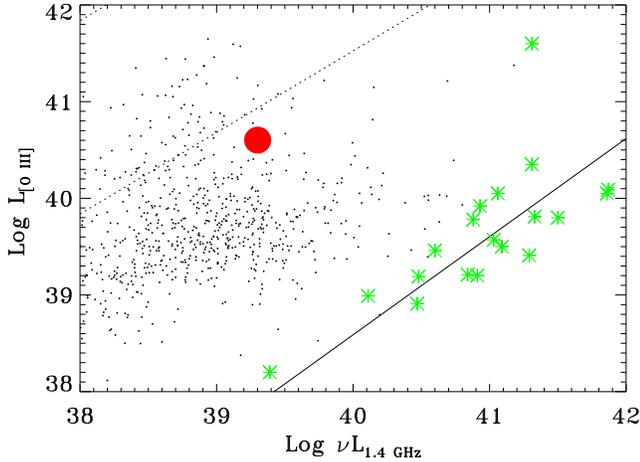}
\caption{Logarithm of the radio vs. [O~III] luminosities (both in erg s$^{-1}$
  units) for the SDSS/NVSS sub-sample ($0.03<z<0.1$) analyzed by
  \citet{baldi09} and mainly composed of FR~0 sources. The solid line
  reproduces the line-radio correlation followed by FR~Is of the 3CR sample
  (green stars). The dashed line marks the boundary of the location of Seyfert
  galaxies (e.g. \citealt{whittle85}). FR~0s show a strong deficit of total
  radio emission, occupying the region to the left of the FR~Is
  \citep{baldi15}. Tol1326-379 (red point) falls into the FR~0s area.}
\label{ropt}
\end{figure}

Summarizing, Tol1326-379 fulfills all the spectro-photometric requirements for
a FR~0 classification listed by \citet{baldi15}. It is an early-type galaxy
with a black hole mass larger than $10^8$ M$_{\odot}$, a low excitation
optical spectrum, and a high radio core dominance. Furthermore Tol1326-379 ,
given its radio luminosity at 1.4 GHz of 2$\times 10^{39}$ erg s$^{-1}$ ,
falls within the region typical of FR~0s in the diagram shown in Fig.\ref{ropt} that compares the
total radio and emission line luminosities.  It also
shows, as characteristic of this class of sources, a large deficit (a factor
of $\sim$300) of radio emission respect to FR~I radio-galaxies with equal
emission line power. Tol1326-379 is also consistent, considering the
relatively large errors related to the line measurement, with the core-[O~III]
luminosity relation found for FR~Is \citep{baldi15} another crucial
prerogative of FR~0s. 

\subsection{Gamma-ray properties}

\begin{figure*}
\centering  
\includegraphics[width=0.99\columnwidth]{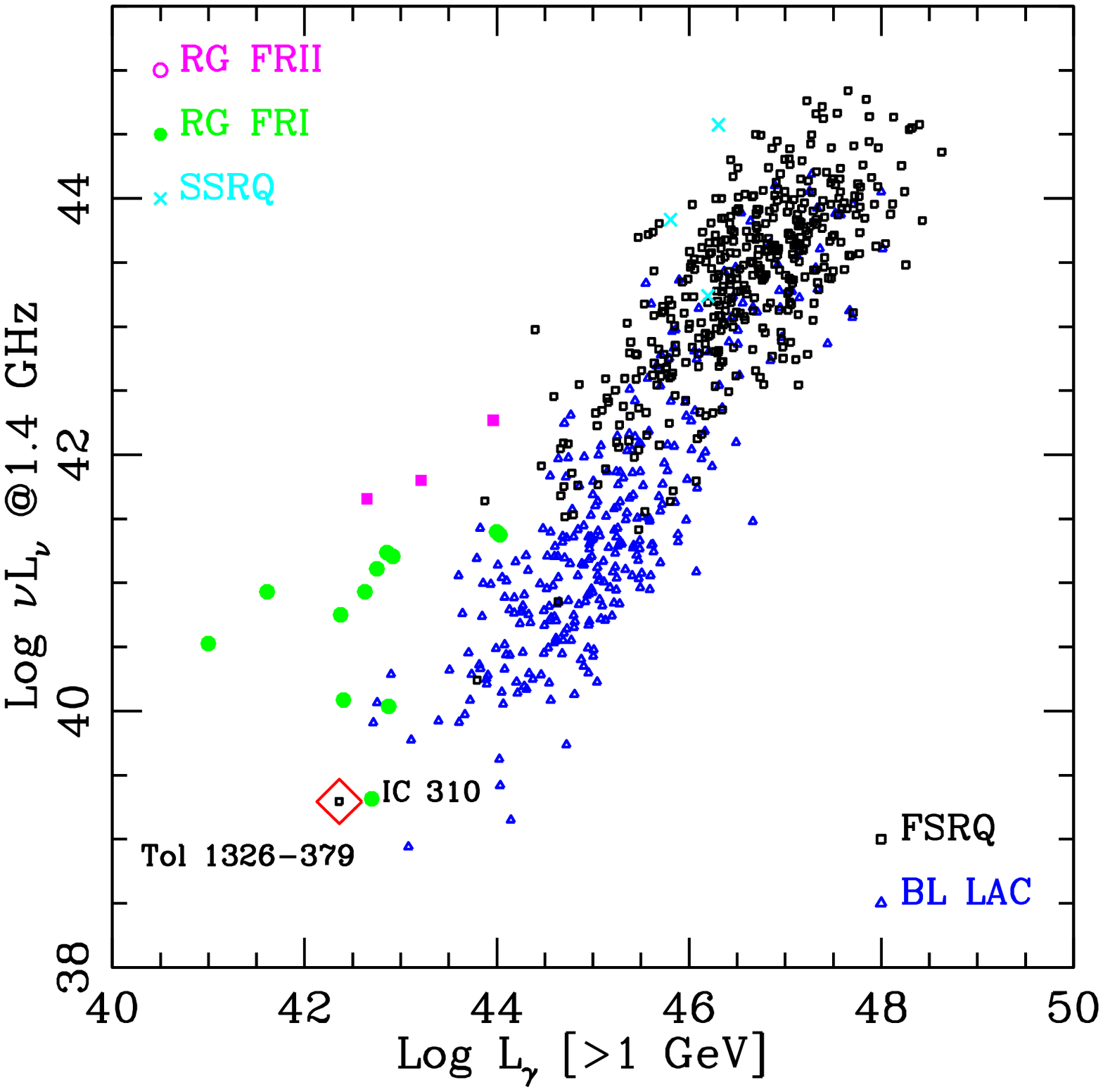}
 \includegraphics[width=0.99\columnwidth]{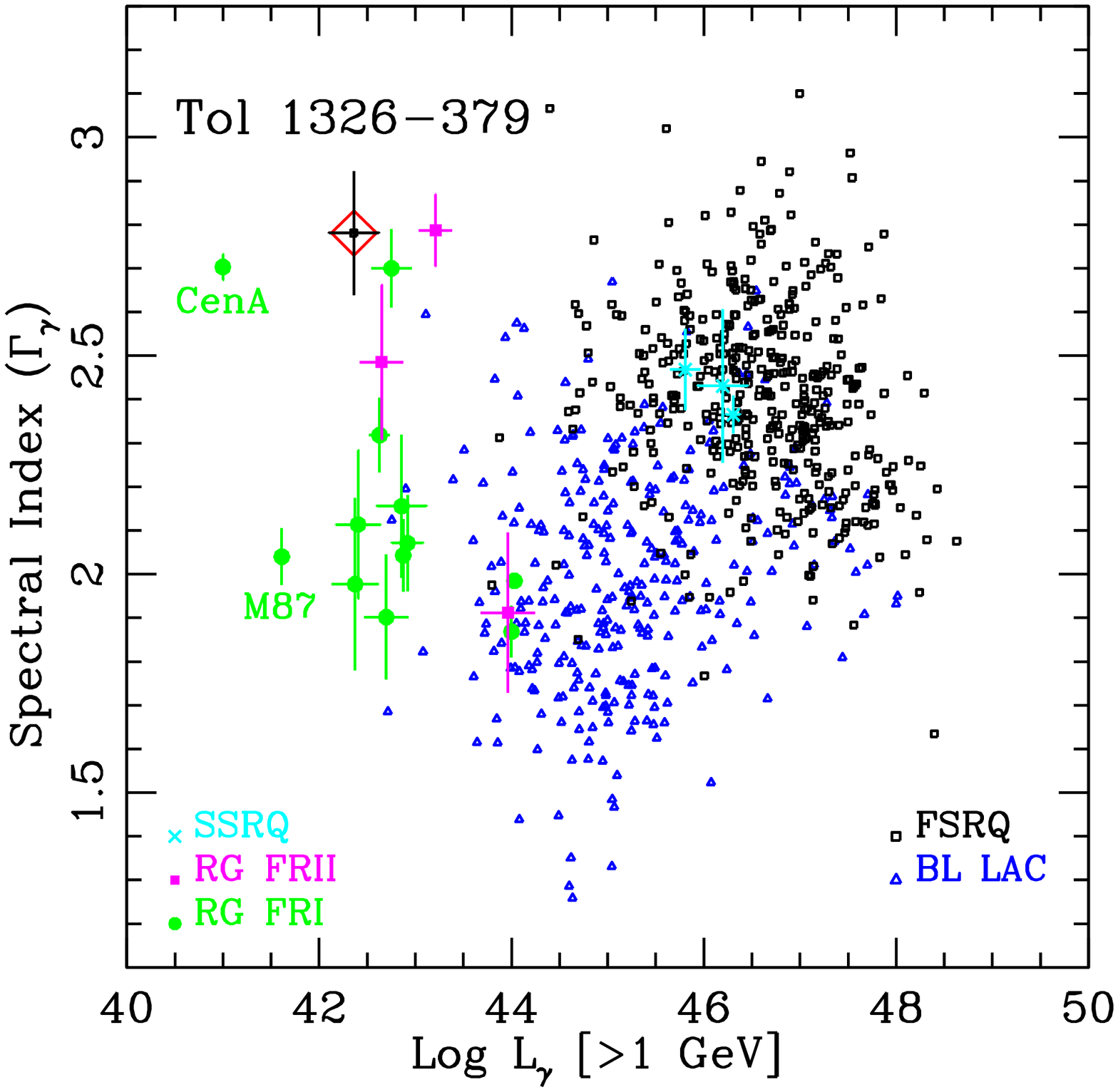}
\caption{Left: the radio luminosity of the 3~LAC FR~I radio galaxies (green
  circles), FR~II radio sources (magenta squares), BL Lacs (open blue
  triangles), FSRQs (open black squares), and SSRQs (cyan crosses) is plotted
  as a function of the $\gamma$-ray luminosity between 1 GeV and 100 GeV.
  MAGN, extended at 1.4 GHz, are more luminous in radio than blazars with
  similar $\gamma$-ray luminosity.  Tol1326-379 (red open diamond) has a
  typical FRI $\gamma$-ray luminosity but falls into the low luminosity tail of
  the blazar strip.  Right: $\gamma$-ray spectral slope versus 1-100 GeV
  luminosity. Tol1326-379 is located into the MAGN region and outside the
  FSRQs zone. For clarity, error-bars are reported for non-blazars only. All
  the luminosities are in erg s$^{-1}$.}
\label{gammar}
\end{figure*}

We explored the position of Tol1326-379 in a plot where the radio luminosity
at 1.4 GHz of all the blazars and MAGN of the 3LAC clean sample (with known
optical classification and redshift) are reported as a function of the
gamma-ray luminosities (see Fig. \ref{gammar}, left panel). We also add
3C~111 and Cen~B (not found in the clean sample because of their low galactic latitudes)
and 3C~120 already reported in the 15 month-sample of MAGN and that has recently
undergone strong flares \citep{casadio15}. As the origin (jet, lobes) of the FornaxA  $\gamma$-ray emission is unclear, we prefer not to include it.

We derived for all sources in the 3LAC catalog the k-corrected 1.4 GHz rest
frame flux density by assuming $\alpha_{\rm r}= 0$ for blazars and 0.8 for
MAGN.\footnote{In some cases the 3LAC radio flux density was provided at
  different radio frequencies (for example at 20 GHz or at 843 MHz). The above
  reported spectral radio slopes were also adopted to convert the listed flux
  to the 1.4 GHz flux.} For  Tol1326-379  we assume the observed value of
$\alpha_{\rm r}=0.37$. As expected, different classes lie in different zones of the
diagram with FSRQs at higher luminosities than BL Lacs.  We recover the well
known radio $\gamma$-ray correlation for blazars \citep{Ghirlanda2010, Ackermann-Radio}.  On
average, MAGN are offset from the blazar strip showing a radio excess with
respect to BL Lacs and FSRQs with similar $\gamma$-ray luminosities. This is due
to the additional contribution from extranuclear radio emission.
Tol1326-379 falls into the low luminosity tail of the blazar strip. It is, together
with IC~310, the least powerful radio galaxy with $\gamma$-ray detection.

To further explore the nature of our source, the $\gamma$-ray spectral slope
($\Gamma$) of blazars and MAGN of the clean 3LAC sample was plotted in
Fig. \ref{gammar}, right panel, as a function of the rest frame isotropic
luminosity above 1 GeV ($L_{>1 {\rm GeV}}$ in ergs cm$^{-1}$ s$^{-1}$). As is
well known, different classes of AGN occupy different locations in the
$\Gamma$-$L_{>1{\rm GeV}}$ plane \citep{abdo10-MAGN}.  FSRQs and SSRQs are in the
upper right part of the diagram (high luminosities and steep spectra), BL~Lacs
in the bottom right side (low luminosities and flat spectra), while RGs are
all at low luminosities, but with a large scatter in spectral indeces. It is
evident that Tol1326-379 falls into the general radio-galaxy area and not into
the FSRQs region.

\section{Discussion}

The 3LAC catalog associates the $\gamma$-ray source 3FGLJ1330.0-3818 to the galaxy Tol1326-379.  
We confirm that no other flat radio source brighter than Tol1326-379 at 5 GHz is present with the 95\% error circle position of 3FGLJ1330.0-3818.
Although flat in the radio, Tol1326-379 is not a FSRQ but a FR~0 radio galaxy, the first source of this new radio class with a $\gamma$-ray counterpart. \\

Several observations support this  conclusion:
\begin{itemize}
\item Tol1326-379 is an early-type galaxy at z=0.02843;
\item it does not show any evidence for broad emission lines, arguing against
its identification as FSRQs;
\item the high values of the line equivalent widths exclude the possibility
  that it is a BL Lac;
\item its line ratios are typical of LEGs;
\item its estimated black hole mass of $M_{\rm BH}=2 \times 10^8$
  M$_{\odot}$ and accretion rate $\dot L=L_{\rm bol}/L_{\rm Edd} \sim
  5 \times 10^{-3}$ are characteristic of radio-loud AGN
  \citep{chiaberge11} associated with low efficiency accretion flows;
\item it is unresolved in the radio images, showing a high core dominance and
  a flat radio spectrum;
\item when put  in a radio versus [O~III] luminosity plot, it falls into the
  FR~0 region and not in the FR~I area.
\end{itemize}

\begin{figure}
\centering  
\includegraphics[width=0.99\columnwidth]{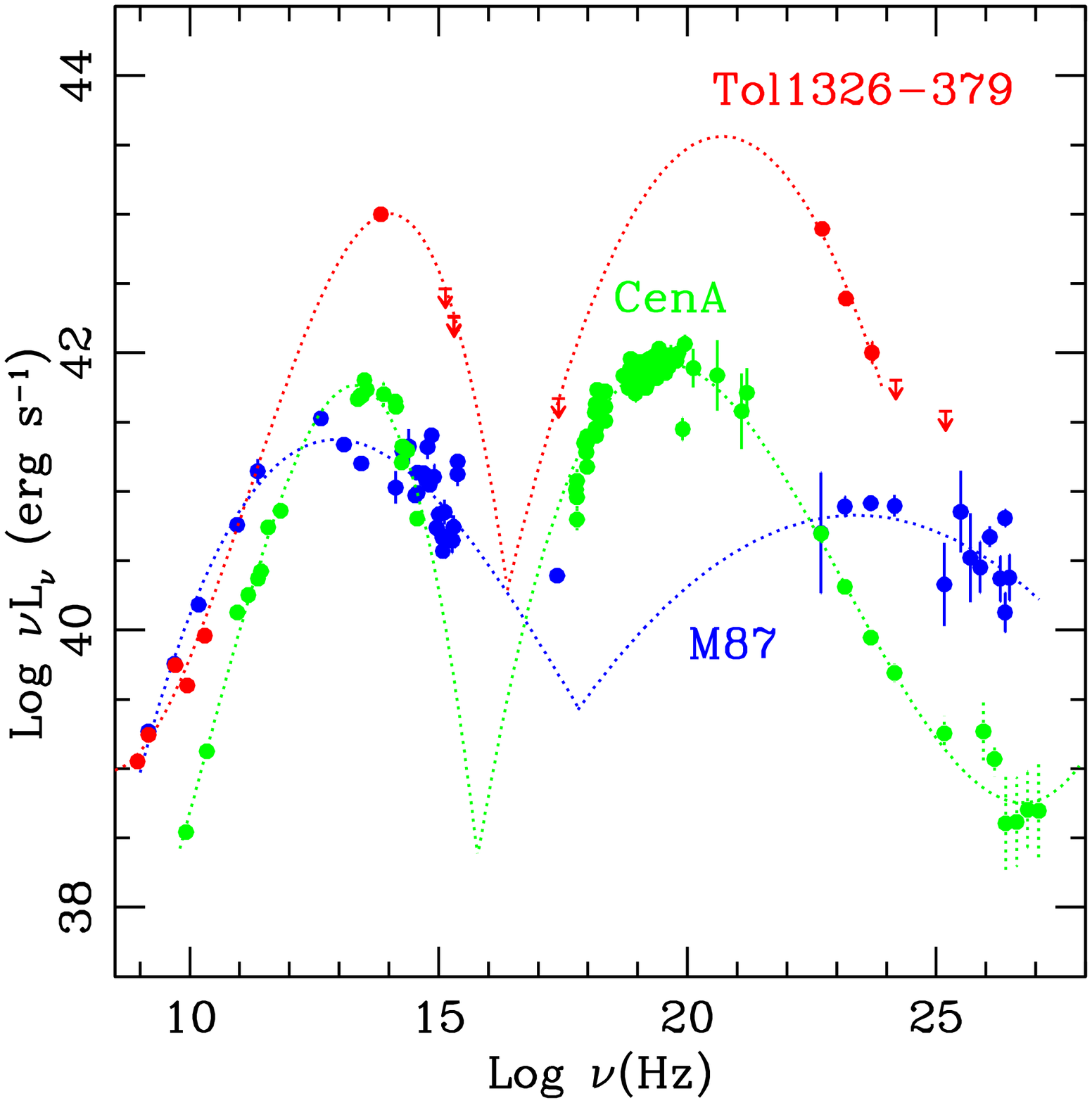}
\caption{SEDs of Tol1326-379 (red symbols) compared to those of Cen~A (green)
  and M~87 (blue). The dotted lines are polynomial functions connecting the
  points and do not represent model fits to data. Data are collected from the
 literature.  M~87: \citet{biretta-M87, despringre-M87, Tan-M87, perlman-M87,
    balmaverde06a, Ackermann-3LAC, broderick15,aharonian-M87,aliu-M87}.
  Cen~A: \citet{meisenheimer-cena, grandi-cena, harmon-cena,
    Ackermann-3LAC,aharonian-cena}. The Compton Gamma Ray Observatory fluxes
  of Cen~A, collected during a multi-frequency campaign in 1995, are provided
  by Steinle H. at the web page http://www.mpe.mpg.de/~hcs/Cen-A/ and included
 in the NASA/IPAC EXTRAGALACTIC DATABASE.}
\label{sed}
\end{figure}

In Fig. \ref{sed}, we present the SED of Tol1326-379. Besides the radio and
Fermi observations already discussed, we complemented it with the WISE, Galex,
and ROSAT data.

Tol1326-379 is detected at high significance by the {\it WISE} satellite in
all four bands and its colors are W1-W2=0.38 $\pm$ 0.03 and
W2-W3=2.58$\pm$0.03, respectively. These are significantly different from
those typical of elliptical galaxies \citep{wright10}. This indicates that the
emission seen, at least, in the W3 band is dominated by the AGN
component. Indeed, Tol1326-379 lies in a region populated by BL~Lac objects
\citep{massaro11}, although offset from the main blazar strip. 

The UV data, from the Galex archive\footnote{http://galex.stsci.edu/GR6/},  are corrected for galactic extinction assuming  E(B-V)=0.0678. 
In the X-ray band, Tol1326-379 was only detected by the PSPC instrument (0.1-2.4 keV) onboard of the ROSAT satellite.
Its count rate reported in the  ROSAT All-Sky Survey Faint Source Catalog\footnote{http://www.xray.mpe.mpg.de/rosat/survey/rass-fsc/} is 0.029$\pm0.013$ c/s.
This can be converted to an unabsorbed flux  of F$_{0.1-2.4~keV}=(7.6\pm 3.4)\times10^{-13}$ erg s$^{-1}$ cm$^{-2}$ if a power law with spectral slope $\Gamma_x=2$ and Galactic absorption of $N_H=5.5\times10^{20}$ cm$^{-2}$ are assumed. 
Due to the low spatial resolution of the UV and X-ray images, and the lack of any spectral
information, the corresponding measurements should be considered as upper
limits, as they include the host galaxy emission. In particular, its soft
X-ray luminosity is compatible with that expected from the hot corona of an
early-type galaxy with the luminosity estimated above for our source
\citep{fabbiano92}.

The SED of Tol1326-379 is compared to those of Centaurus A and M87, the
prototype nearby FR~I radio galaxies.  Despite of the similar radio luminosity, M~87 is less luminous than Tol1326-379 by a factor of 30 at 1 GeV and has a
flat SED in the $\gamma$-ray domain. On the contrary, Centaurus A and
Tol1326-379 are quite similar in shape but the former source is 
about two orders of magnitude fainter.

\citet{balmaverde06b} found that the SEDs of the FR~I nuclei differ from those
of BL~Lacs. This is partly related to the frequency shift in the SED due to
the relativistic beaming, but also to differences in the emitting
regions. Indeed, due to the presence of a velocity stratification with the
relativistic jets \citep{kovalev2007,nagai2014} in BL~Lacs we are seeing the
regions of the jet with the highest Doppler factor (the so-called jet spine)
while in FR~Is the emission could be dominated by the slower jet layers. The SED
differences might be witnessing the diverse physical conditions in these two
regions \citep{chiaberge00c,ghisellini05,tavecchio2014} leading to a
dependence of their shape on the viewing angle.

This effect might account for the SED differences between M~87 and
Tol1326-379 and, on the other hand, the similarities with Centaurus~A. Indeed,
M~87 jet is seen at a rather small angle from the line of sight with
$\theta\sim 15-25^{\circ}$ \citep{acciari2009}, while Cen~A lies close to the
plane of the sky, $\theta\sim50-80^{\circ}$ \citep{Tingay1998}. We argue that
the SED shapes are, similarly to what is derived from the comparison between
BL~Lacs and FR~Is, mainly driven by the jet viewing angle. Indeed, by
excluding Centaurus~A,  all  the FR~Is have flat $\gamma$-ray spectra and are seen at small angles (see Table \ref{tab:data}).
The only exception is 3C~120, that has a slope comparable to that of Cen~A (and Tol1326-379), but  smaller inclination angle.
3C~120 is, however,  a peculiar radio galaxy. Although classified as FR~I, it  has optical-UV-X-ray properties typical of FR~IIs.
It shows broad optical emission lines \citep{tadhunter}, an UV bump, and an K$_{\alpha}$ iron line in the X-ray spectrum \citep{zdziarski, ogle, kataoka, Chattarjee},  all clear signatures of the presence of an efficient accretion disk.
Indeed, in  Fig. \ref{gammar} (right panel) 3C~120 falls between 3C~111 ($\Gamma$=2.79,L$_{\gamma}$=43.21), and Pictor~A ($\Gamma$=2.49,L$_{\gamma}$=42.65), two (FR~II)  BLRGs detected by {\it Fermi}.

Although we do not have any direct estimate for the orientation of
Tol1326-379, we speculate that it is oriented at a large angle and this causes
the similarity between its SED and that of Cen~A. 

Alternatively, if we are seeing the jet of Tol1326-379 at a small angle, its SED would be intrinsically
different from those of the FR~Is. This would be the first indication of a
discrepant property between these two classes (other than the paucity of
extended radio emission in FR~0) and might be an important clue to understand
their nature. Interesting enough, the Compton peak in Tol1326-379 appears to be more
prominent than the Synchrotron one. This generally occurs in FSRQs, where a
surplus of seed photons coming from the accretion disk, the broad line region
and the torus, contributes to the high energy emission by External Compton
\citep{sikora2009}.  Incidentally we note that Tol1326-379 is
characterized by a steep $\gamma$-ray spectrum
($\Gamma_{\gamma}=2.78\pm0.14$), more similar to that generally observed in FSRQs (and 
their misaligned population, i.e BLRGs and SSRQs) than in BL Lacs (and FRIs).
It is, however, improbable that an External Compton mechanism is responsible for
the cooling of the jet particles of Tol1326-379.  Its nuclear environment is
poor in photons as indicated by the low accretion rate.  
It is then more
plausible that its $\gamma$-ray luminosity is sustained by different jet
components that mutually interact amplifying the IC emission as suggested for
FR~Is.  The excess of $\gamma$-ray radiation could then reflect different
physical conditions of the high energy dissipation regions (i.e. of the spine
and/or the layers).  

Progress in our understanding of the properties of Tol1326-379 can come from a
better definition of its SED. In particular, the information in the X-rays can
be improved with a firm detection and with a measurement of the spectral slope
in this band. This might be used to test the indication that the Compton peak
in Tol1326-379 is more prominent than the synchrotron one.

Since a large diversity of spectral behavior among
this class of sources, as already observed in blazars, cannot be excluded, 
other observations of $\gamma$-rays emitting FR~0s are necessary to consolidate the overall picture.

\begin{table}
\caption{Jet inclination angle from literature (Ref.) , $\gamma$-rays spectral slope ($\Gamma$)  and luminosity $L_{\gamma}$ [$>1$ GeV] (erg s$^{-1}$) of the FR~I radio
  galaxies. }
\begin{tabular}{lllll}
\hline
Source & $\theta$  & $\Gamma$ &L$_{\gamma}$ &Ref.\\
\hline
3C~78        & $~50  ^{\circ}$  & 2.07$\pm$0.11 & 42.92 & 1\\
IC~310       & $<$20$^{\circ}$    & 1.90$\pm$0.14 & 42.70&2\\
NGC1275      & 30-$55^{\circ}$  & 1.95$\pm$0.01 &44.03 &3\\
B20331$+$39  & $<$45$^{\circ}$    & 2.11$\pm$0.17 &42.40&4\\
3C~120$^{a}$            &$18-22^{\circ}$ &2.7$\pm0.1$&42.75& 5\\
PKS~0625-45  & $<$61$^{\circ}$ & 1.87$\pm$0.05 &44.00 &6\\
3C~189       & $~27^{\circ}$  & 2.16$\pm$0.16 &42.86 &7\\
3C~264       & $\sim$50$^{\circ}$ & 1.98$\pm$0.19 &42.37 &8\\
M~87         & 15-$25^{\circ}$  & 2.05$\pm$0.06 &41.61&9\\
Cen~A         & 50-$80^{\circ}$  & 2.70$\pm$0.03 &41.00& 10\\
Cen~B$^{a,b}$      & $<80^{\circ}$  &      2.32$\pm$0.09&42.63&11\\
NGC~6251     & 10-$40^{\circ}$  & 2.04$\pm$0.08 &42.87& 12\\
\hline
\end{tabular}

{\footnotesize
$^a$ -- Cen~B and 3C~120 are not  in the clean 3LAC sample.\\
$^b$ --  Cen~B: Jet inclination upper limit estimated using  the jet/counterjet flux ratio $J \sim 2$ at 4.8 GHz provided by \cite{jones}. Possible $\gamma$-ray contribution from the lobes \citep{cenB}.}

{\scriptsize
(1) \cite{3C78-i}, (2) \cite{IC310-i}, (3) \cite{NGC1275-i}, (4) \cite{giovannini2001}, (5) \cite{jorstad}, (6) \cite{venturi2000}, (7) \cite{3C189-i}, (8) \cite{3C264-i}, (9) \cite{acciari2009}, (10) \cite{Tingay1998}, (11) \cite{jones}, (12)  \cite{NGC6251-i}.}
\label{tab:data}
\end{table}
 
\section*{Acknowledgments}
We are grateful to F. Paresce for carefully reading the paper and useful comments which improved the final version. 
We thank the referee for constructive comments/suggestions on the manuscript. 
This research made use of the NASA/ IPAC Infrared Science Archive and
Extragalactic Database (NED), which are operated by the Jet Propulsion
Laboratory, California Institute of Technology, under contract with the
National Aeronautics and Space Administration.
Part of this work is based on archival data,  software or online services provided by the Italian Space Agency (ASI) Scientific Data Center (ASDC).
\bibliography{./my_r}

\end{document}